\documentclass[12pt, a4paper]{jpconf}

\newcommand{\al}{\alpha}

\newcommand{\ben}{\begin{eqnarray}}
\newcommand{\een}{\end{eqnarray}}
\newcommand{\be}{\begin{equation}}
\newcommand{\ee}{\end{equation}}
\newcommand{\ba}{\begin{eqnarray}}
\newcommand{\ea}{\end{eqnarray}}
\newcommand{\n}{\label}
\newcommand{\no}{\noindent}
\newcommand{\la}{\lambda}
\newcommand{\ga}{\gamma}
\newcommand{\ro}{\rho}

\begin{document}

\title{k-essence and extended tachyons in brane-worlds }

\author{Luis P. Chimento $^1{^,}^2$ and Mart\'{\i}n G. Richarte $^2$}
\address{$^1$ Department of Theoretical Physics, University of the Basque Country,
P.O. Box 644, 48080 Bilbao, Spain and IKERBASQUE, the Basque Foundation for Science, 48011, Bilbao, Spain.}

\address{$^2$ Departamento de F\'{\i}sica, Facultad de Ciencias Exactas y
Naturales,  Universidad de Buenos Aires, Ciudad
Universitaria, Pabell\'on I, 1428 Buenos Aires, Argentina}

\ead{chimento@df.uba.ar, martin@df.uba.ar }

\begin{abstract}
We study a k-essence field evolving linearly with the cosmic time and the atypical k-essence model on a homogeneous and isotropic flat 3-brane. We show that the k-field is driven by an inverse quadratic polynomial potential. The solutions represent expanding, contracting or bouncing universes with a finite time span and some of them end in a big crunch or a big rip. Besides, by selecting the extended tachyonic kinetic functions we analyze the high and low energy limits of our model, obtaining the nearly power law solution. We introduce a tachyon field with negative energy density and show that the universe evolves between two singularities.  
\end{abstract}


\section{Introduction}

Recent progresses in Superstring/M-theory have offered  new perspectives to understand the cosmological evolution of the universe \cite{ADD1}, \cite{RS}. In these theories our Universe is conceived as a 3-brane embedded in a higher dimensional spacetime usually referred as the bulk \cite{RM}. Among many interesting models coming from this new  scenarios, there are two possibilities that seem to resolve different problems in the standard cosmology, the RS \cite{RS} and DGP  \cite{DGP1} scenarios.

Several works have been investigated in cosmological scenarios with non-canonical kinetic term usually known as k-essence models \cite{INFLKE}-\cite{10}. In particular, some efforts in the framework of k-essence have been directed toward model building using power law solutions which preserve \cite{6}, \cite{PAD1},\cite{AFE1}, \cite{CHFE1} or violate the weak-energy condition \cite{ACHR}. In addition, a k-essence model with a divergent sound speed, called atypical k-essence, was carefully analyzed in  Refs. \cite{8},  \cite{MCL}. In the latter case it showed that the model fix the form of the Lagrangian for k-essence matter. Then it would be interesting to study brane-worlds models supported by k-essence. 

Recently, power law solutions on a 3-dimensional brane  coupled  to the tachyonic field  were obtained \cite{SA1} by using a well known  algorithm developed in \cite{PAD1}. Further the high and low energy limits for the tachyonic potential turns out to be $V= V_{0}\phi^{-1}$  and $V= V_{0}\phi^{-2}$ respectively. 

Here, we  present a model with k-essence localized on the Friedmann-Robertson-Walker (FRW) brane and obtain the scale factor and potential for a k-field evolving linearly with the time. Also, we examine the divergent sound speed model and apply these results to the extended tachyon field cases. Finally the conclusions are stated.

\section{k-essence in brane-worlds}
We will explore the evolution of a universe filled with a k-essence field $\phi$ in a flat FRW spacetime. Using the perfect fluid analogy, the energy density and pressure are given by
\be
\n{ro}
\rho_{\phi}=V(\phi)[F-2xF_{x}], \qquad  p_{\phi}={\cal{ L}}=-V(\phi)F(x).
\ee
where $F(x)$ is a function of the kinetic energy $x=-\dot{\phi}^{2}$, $F_{x}=d\,F/d\,x $  and $V(\phi)$ is a positive  potential.
We shall focus on cosmological branes  with the induced metric $g_{\mu\nu}={\mbox{diag}} [-1,a^{2}\delta{j_i}]$, where $a$ is the scale factor.  Then the modified Einstein equations on the brane are \cite{BRA}
\ben
3H^{2}=\rho_{\phi}+ \frac{3}{\lambda^{2}}\rho_{\phi}^2 \label{fm}
,\\
\big(F_x+2xF_{xx}\big)\ddot\phi+3HF_{x}\dot\phi+\frac{V'}{2V}\big(F-2xF_{x}\big)=0,\label{cphi1}
\een
with $H=\dot{a}/a$ the Hubble expansion rate and $'\equiv d/d\phi$. Also, by writing the equation of state  for the k-essence as $p_{\phi}=(\gamma_{\phi}-1)\rho_{\phi}$, the barotropic index reads $\gamma_{\phi}=-2xF_{x}/(F-2xF_{x})$. The quadratic term (\ref{fm}) modifies the standard cosmology dominating at the early times, $H \propto \rho_{\phi}/\la$, whereas in the limit $\la \rightarrow \infty$ we recover standard relativity $H\propto \sqrt{\rho_{\phi} }$.

For $x=x_0=const$ and $F_x(x_0)=0$, we get $\gamma_{\phi}=0$. Hence, from Eqs. (\ref{fm}) and (\ref{cphi1}) the energy density and the expansion rate become constants and the brane exhibits a de Sitter phase. 

Other kind of solutions can be found by re-writing the Eqs. (\ref{fm})-(\ref{cphi1})
\be
3H^{2}=V{\al}+ \frac{3}{\lambda^2}{\al}^2V^{2}\label{fm2}
,\qquad 
\frac{\dot{\al}}{{\al}}+\frac{\dot{V}}{V} +3H\gamma_{\phi}=0\label{cphi2},
 \ee
with $\al=\rho_{\phi}/V=F-2xF_x$ and assuming the constraint $\al=\al_{0}=const$. It is satisfied for any kinetic function $F$ when $i.$ the k-field evolves linearly with the cosmological time, $\phi=\phi_{0}t$ with $x_0=-\phi_{0}^2=const$ or $ii.$ for the kinetic function fulfilling $F^\infty-2xF^\infty_{x}={\al}_{0}$ \cite{6}, \cite{8}, \cite{CHFE1}, hence 
\begin{equation} 
\label{fdiv}
F^\infty={\al}_{0}  + \beta\sqrt{-x},
\end{equation}
where $\beta$ is an arbitrary constant. It is associated with a divergent sound velocity and with the extended tachyon model considered in \cite{6}. The function $F^\infty$ generates a  divergent sound speed model called atypical k-essence \cite{8}, \cite{MCL}. 

\vskip .5cm

\no $i.$ In this case the barotropic index $\ga_\phi=\ga_0=const$ and one finds the potential $V=V_{0}a^{-3\gamma_{0}}$, after integrating the conservation equation (\ref{cphi2}b). For this potential the Eq. (\ref{fm2}a) reads
\be
\label{fmadet}
{\dot{y}}^{2}=3\ga_0^2\left[\ro_{0}y +\frac{3\ro^{2}_{0}}{\la^{2}}\right],
\ee
with $y=a^{3\gamma_{0}}$ and  $\ro_{0}=V_{0}{\al}_{0}$. Its general solution is given by
\be
\label{adet}
a(\tau)=(3\ro_0)^{1/3\gamma_{0}}\left[\frac{\tau^2}{4}\pm\frac{\tau}{\la}\right]^{1/3\gamma_{0}},
\ee
where $\tau=\gamma_{0}t$. This solution has singularities at $\tau_s=0$ and/or  $\tau_s=\mp 4/\la$. For $\ro_0>0$, we have four expanding universes, two of them with $\gamma_{0}>0$ evolve from an initial singularity at $\tau_s=0$ or $\tau_s=4/\la$.  The scale factor begins as $a\propto (\tau-\tau_s)^{1/3\gamma_{0}}$ in the high energy regime and ends as $a\propto \tau^{2/3\gamma_{0}}$ in the low energy regime. The remaining two universes with $\ga_0<0$ end in a final big rip at $\tau_s=0$ or $\tau_s=-4/\la$ having the final behavior $(\tau_s-\tau)^{1/3\ga_0}$. For $\ro_0<0$, the solutions have an extremum at $\tau_{e}=2/\la$ where $a_e=(-3\ro_0/\la^2)^{1/3\ga_0}$. They represent two universes with a finite time span, one of them begins from a singularity, reaches a maximum value $a_e$, and ends in a big crunch at $\tau_s=4/\la$. The other begins from a singularity at $\tau_s=0$, reaches a minimum at $a_e$, where it bounces, and ends in a final big rip at $\tau_s=4/\la$. In addition we have the time reversal of these solutions. 

In turn, by combining $\phi(t)$ with $a(t)$ and $V=V_{0}a^{-3\gamma_{0}}$, we obtain  the following potential 
\begin{equation}
\label{Vphi}
V=\frac{V_0}{3\ro_0}\left[\frac{\ga_0^2 \phi^2}{4\phi_0^2}\pm\frac{\ga_0\phi}{\la\phi_0}\right]^{-1}.
\end{equation}
In the early universe quadratic contributions in $\rho_{\phi}$ become important and the potential approaches $V\propto \pm\la \phi^{-1}$ when $\la\rightarrow 0$. It seems to be the counterpart of the  solution for quintessence cosmology driven by an inverse square potential \cite{KIEM}, \cite{NOS}. In the limit $\la\rightarrow \infty$, the potential $V\propto \phi^{-2}$ and we recover the power law solution  \cite{PAD1}, \cite{AFE1} \cite{CHFE1}. 

The Eq. (\ref{Vphi}) shows that the brane correction shifts the inverse square potential to the inverse linear one at high energy. It has a minimum at $\phi_{e}=2\phi_0/\la\ga_0$ or a maximum at $\phi_{e}=-2\phi_0/\la\ga_0$. 
Finally, by associating the brane equations to an effective fluid description, we find a linear barotropic equation of state in the low energy limit. However, in the high energy  regime the k-essence source becomes a modified Chaplygin gas, $P_{ef}\approx (2\gamma_{0}-1)\ro_{ef}+\pm\gamma_{0}\la\ro^{-1/2}_{ef}/\sqrt{3}$.

\vskip .5cm

\no $ii.$ By combining the Eq. (\ref{fdiv}) and its associated  barotropic index $\gamma_{\phi}=-\beta\sqrt{-x}/\al_0$ with the Eq. (\ref{cphi2}b) we find  $H=\al_0 V'/(3\beta V)$. So by using $\rho_{a}=\al_0 V(\phi)$ into the Eq. (\ref{fm}) one gets the potential
\begin{equation}
\label{Vaty}
V^\infty=\frac{1}{3}\left[\frac{\beta}{4\al_0}\phi^{2}\pm \frac{\phi\beta}{\la}\right]^{-1}.
\end{equation}
The potential on the brane (\ref{Vaty}) generalizes the inverse square one of the Friedmann cosmology shifting the divergence to $\phi_{\infty}=\pm4\al_0/\la\beta$. So, for asymptotic power-law expansions the linear k-field model driven by a nearly inverse square potential and the atypical model are isomorphic.

\subsection{Extended Tachyons in brane-worlds}

Recently, it was proposed that the tachyon Lagrangian could be extended  in such a way to allow the barotropic index any value \cite{6} generating phantom and complementary tachyons in addition to the ordinary one \cite{6}, \cite{CFKR}. This extended tachyons generalize the Chaplygin gas introducing the phantom and complementary Chaplygin gases. The kinetic functions for these tachyons are
\begin{eqnarray}
\n{to}
F_{t}=(1-\dot{\phi}^{2r}_{t})^{1/2r},   \qquad   0<\ga_{t}=\dot{\phi}^{2r}_{t}<1 
\\
\n{tp}
F_{ph}=(1+\dot{\phi}^{2r}_{ph})^{1/2r},   \quad   -\infty<\ga_{ph}=-\dot{\phi}^{2r}_{ph}<0 
\\
\n{ft}
F_{c}=-(\dot{\phi}^{2r}_{c}-1)^{1/2r},   \quad   1<\ga_{c}=\dot{\phi}^{2r}_{c}<\infty
\end{eqnarray}
where $r$ is a real parameter. The scale factor and potential for the three sets of extended tachyon fields are given by Eqs. (\ref{adet}) and (\ref{Vphi}) with $\gamma_{0}=\phi^{2r}_{0t}$, $\gamma_{0}=-\phi^{2r}_{0ph}$, $\gamma_{0}=\phi^{2r}_{0c}$ and ${\al}_{0}=(1-\phi^{2r}_{0t})^{-1/2}$, ${\al}_{0}=(1+\phi^{2r}_{0ph})^{-1/2}$, ${\al}_{0}=-(\phi^{2r}_{0c}-1)^{-1/2}$ respectively. 

The ordinary and complementary tachyons lead to expanding scenarios while the former has an accelerated expansion for $\phi^{2r}_{0t}<2/3$. A universe dominated by a phantom tachyon ends in a big rip at $\tau_s=0$ or $\tau_s=-4/\la$ where the scale factor blows up as $a\propto \tau^{-1/3\phi^{2r}_{0ph}}$  and  the potential diverges as $V\propto \phi^{-1}_{ph}$. The  ordinary and the complementary tachyons satisfy the weak energy condition $\ro\geq0$ and $\ro+p\geq0$, and the null energy condition $\ro+p\geq0$ while the phantom tachyon violates both conditions. 

As Eq. (\ref{fm}) is quadratic in the energy density, we introduce a tachyon with negative energy density by making $F\to -F$ in Eqs. (\ref{to})-(\ref{ft}) and keeping $V>0$. The universe evolves between two singularities having a finite time span $t_{s}=4/\la|\gamma_{0}|$. The scale factor has a maximum for ordinary and complementary tachyons, ending in a big crunch or has a minimum where the phantom tachyon bounces before  it blows up in a big rip.  
 
\section{Summary}

We have found  solutions for 3-dimensional cosmological brane with k-essence and extended tachyons, whose metric  corrpesponds to a  spatially flat FRW spacetime,  when the ratio $\ro/V$ is constrained to be constant for linear k-field or for the atypical k-essence model. For $\ro_{0}>0$ we have obtained power law behavior at the initial and final stages. However, in the case $\ro_{0}<0$ the universe bounces, has a finite time span and ends in a big crunch ($\ga_0>0$) or a big rip ($\ga_0<0$). The quadratic brane correction has shifted the inverse square potential to the inverse linear one at high energy, so that the k-field is driven by an inverse quadratic polynomial potential. Finally, we have analyzed the extended tachyons with negative energy density and showed that the scale factor evolves between two singularities having a finite time span which depends on the brane tension and the barotropic index. For the ordinary and complementary tachyons the scale factor ends in a big crunch while for the phantom one it bounces and ends in a big rip.

\ack
LPC thanks the University of Buenos Aires for the partial support of this work during its different stages under Project X044, and the Consejo Nacional de Investigaciones
Cient\'{\i}ficas y T\'ecnicas under Project PIP 114-200801-00328.  MGR is supported by  the Consejo Nacional de Investigaciones Cient\'{\i}ficas y T\'ecnicas (CONICET).

\end{document}